\documentclass{JINST}

\usepackage{multirow}
\usepackage{epsfig}
\usepackage{url}
\usepackage{subfigure}
\usepackage{amsmath}

\preprint{arXiv:1011.0337\\
SLAC-PUB-14255\\
DESY 10-146}

\title{Results from a prototype chicane-based\\
energy spectrometer for a Linear Collider}

\author{A.~Lyapin$^{a,d}$\thanks{Corresponding author}, H.J.~Schreiber$^b$, M.~Viti$^b$, C.~Adolphsen$^c$, R.~Arnold$^c$, S.~Boogert$^d$, G.~Boorman$^d$, M.~V.~Chistiakova$^e$, F.~Gournaris$^a$, V.~Duginov$^f$, C.~Hast$^c$, M.~D.~Hildreth$^g$, C.~Hlaing$^e$, F.~Jackson$^h$, O.~Khainovsky$^e$, Yu.~G.~Kolomensky$^e$, S.~Kostromin$^f$, K.~Kumar$^{c,i}$, B.~Maiheu$^a$, D.~McCormick$^c$, D.~J.~Miller$^a$, N.~Morozov$^f$, T.~Orimoto$^{e,j}$, E.~Petigura$^e$, M.~Sadre-Bazzaz$^e$, M.~Slater$^k$, Z.~Szalata$^c$, M.~Thomson$^k$, D.~Ward$^k$, M.~Wendt$^l$, M.~Wing$^a$ and M.~Woods$^c$\\
\llap{$^a$}University College London, London, UK,\\ 
\llap{$^b$}Deutsches Electronen Synchrotron DESY, Hamburg and Zeuthen, Germany,\\
\llap{$^c$}SLAC National Accelerator Laboratory, Menlo Park, California, USA,\\
\llap{$^d$}Royal Holloway, University of London, Egham, UK,\\ 
\llap{$^e$}University of California and Lawrence Berkeley National Laboratory, Berkeley, California, USA,\\ 
\llap{$^f$}Joint Institute for Nuclear Research, Dubna, Moscow Region, Russia,\\
\llap{$^g$}University of Notre Dame, Notre Dame, Indiana, USA,\\
\llap{$^h$}Daresbury Laboratory, Daresbury, UK,\\ 
\llap{$^i$}University of Massachusetts, Amherst, Massachusetts, USA\\
\llap{$^j$}California Institute of Technology, Pasadena, California, USA,\\
\llap{$^k$}University of Cambridge, Cambridge, UK,\\ 
\llap{$^l$}Fermi National Accelerator Laboratory, Batavia, Illinois, USA,\\
  E-mail: \email{Alexey.Lyapin@rhul.ac.uk}
}

\abstract{The International Linear Collider (ILC) and other proposed high energy $e^+e^-$ machines aim to measure
with unprecedented precision Standard Model quantities and new, not yet discovered phenomena.
One of the main requirements for achieving this goal is a measurement of the incident beam energy with an uncertainty close to $10^{-4}$.
This article presents the analysis of data from a prototype energy spectrometer commissioned in 2006--2007 in SLAC's End Station A beamline.
The prototype was a 4-magnet chicane equipped with beam position monitors measuring small changes of the beam orbit through the chicane at different beam energies.
A single bunch energy resolution close to ${5\cdot 10^{-4}}$ was measured, which is satisfactory for most scenarios.
We also report on the operational experience with the chicane-based spectrometer and suggest ways of improving its performance.}


\begin{document}

\section{Introduction}

The physics potential of the next $e^+e^-$ Linear Collider
depends greatly on precision energy measurements of the electron
and positron beams at the interaction point (IP).
Beam energy measurements are mandatory for the precision determination of the fundamental properties of particles created in the processes of interest.
For example, measuring the top mass to order of $100-200$~MeV or measuring the mass of the Standard Model Higgs boson to about 50~MeV using the Higgs-strahlung process requires the luminosity-weighted
collision energy to be known to a level of ${(1-2)\cdot 10^{-4}}$ 
to avoid
this being the dominant uncertainty~\cite{ref:ILCReportPhysics}.
The strategy proposed in the International Linear Collider (ILC) design report~\cite{ref:ILCReportAccelerator}
is to have redundant beam-based measurements capable of achieving a $10^{-4}$
relative precision on a single beam, which would be available in real time
as a diagnostic tool to the operators. Also, physics reference channels, such as
$e^+ e^- \rightarrow \mu^+ \mu^- \gamma$, where the muons are resonant with
the known Z-mass, are expected to provide valuable cross-checks of the
collision energy scale, but only long after the data had been recorded.

The primary method planned to perform the beam energy ($E_b$) measurements at the ILC is a non-invasive energy spectrometer using beam position monitors (BPMs).
 The proposed setup is similar 
 to that used for calibrating the energy scale for the W-mass measurement 
 at LEP-II \cite{ref:LEP-II}. At the ILC,
however, the parameters of the spectrometer are tightly constrained to provide
limited emittance dilution at the highest ILC energy $E_b = 500$~GeV.

Initially, a 3-magnet chicane located upstream of the interaction point 
just after the energy collimators of the beam delivery system (BDS) 
was proposed~\cite{ref:lcdet04-31}.
However, the baseline ILC spectrometer design uses two dipole magnets to produce
a beam displacement $x$, while two more magnets return the beam to the nominal
beam orbit.
 For such a chicane, the beam energy (to first order) is then given by

\begin{equation}
E_b = \frac{c \cdot e \cdot L}{x} \int\limits_{\mathrm{magnet}} B \, \mathrm{d}l  \hspace{1mm},
\label{eqn:energy}
\end{equation}

\noindent
where $L$ is the distance between the first two magnets,
 $\int{B \, \mathrm{d}l}$ the integral of the magnetic field in each magnet,
 $c$ the speed of light and $e$ the electric charge of the electron.

The 4-magnet chicane
avoids spurious beam displacement signals in the BPMs due to the inclination of the beam trajectory, 
and thus systematic errors in $E_b$ measurements. For this reason,
a 4-magnet spectrometer, which maintains the beam axially with respect
to the axis of the cavity BPMs, seems preferable to a more conventional
3-magnet chicane. 
In both cases the magnetic field in the spectrometer chicane can be recorded
and reversed for studying systematic effects
 without changing the beam direction downstream of the spectrometer \cite{ref:specandpol}.


A dispersion of 5~mm at the centre of the chicane can be introduced routinely without a significant degradation of the beam emittance due to synchrotron radiation. When operating with a fixed dispersion of 5~mm 
over the whole energy range, a micrometre-level BPM resolution is needed. This resolution can be achieved with
cavity BPMs \cite{ref:BPM}. Since the spectrometer bending magnets
need to operate at low fields when running the ILC at the Z-pole,
the magnetic field measurement may not be accurate enough to provide the required level of precision.
A significantly improved BPM resolution would, however, allow the magnets to be run at the same field for both the Z-pole and highest energy operation.


Some original energy resolution studies of the SLAC prototype 4-magnet chicane 
were presented in reference~\cite{ref:Viti}. The analysis used 
calibrated beam position readings but revealed that due to
small differences between the magnets in the chicane the beam inclination
also needs to be considered. The analysis has here been
 extended by using complex BPM readings that contain the information
on both the beam offset and inclination. This approach eliminates the need
for position calibration of the BPMs, while the whole system can be calibrated
by means of an energy scan.


In this publication we 
 estimate the resolution of the spectrometer to compare it 
with the result of ${8.5\cdot 10^{-4}}$ measured in \cite{ref:Viti}.
We also consider the impact of different systematics on the energy measurement
in order to improve the resolution to the $10^{-4}$ level in future experiments.

\section {Test beam setup and spectrometer hardware configuration}

A prototype test setup for a 4-magnet chicane was commissioned
in 2006 (the T-474 experiment)
and extended in 2007 (the T-491 experiment) in the End Station A (ESA) beamline
at the SLAC National Accelerator Laboratory~\cite{ref:T-474/491}.

In our experiments the electron beam generated by the main Linear Accelerator at SLAC
was transported to the ESA experimental area through the 300~m long A-line, which includes bending and focusing magnets, diagnostic instruments, such as stripline and Radio Frequency (RF) cavity BPMs, charge sensitive toroids, a synchrotron light monitor, 
profile screens and waveguide pick-ups.
The SLAC linac provided single bunches at 10 Hz and a nominal energy of
 28.5~GeV, 
a bunch charge of ${1.6\cdot10^{10}}$ electrons, a bunch length of 500 $\mu$m 
and an energy spread of 0.15\%, i.e.
 a beam with properties similar to the ILC expectations
 at the highest energy currently available for electrons.

These beam parameters allowed us to test
the capabilities of the proposed spectrometer under realistic conditions.
Two feedback systems were in place for the ESA beam: 
one for its position and one for the energy. The position feedback stabilised the beam position
and angle using cavity BPMs and corrector magnets upstream of the ESA area. 
The energy feedback stabilised the energy by controlling the phase
 of the klystrons, and thus the accelerating gradient,
 in one of the linac sections.
 The energy feedback was also used for offsetting the energy from the nominal value in approximately 50~MeV steps within a $\pm$100~MeV range, thus providing a rough energy calibration for the spectrometer.

Remaining beam energy drifts change the beam orbit through the transfer line, resulting in increased beam losses as the trajectory wanders off the optimal one. Monitoring these losses and correcting for the drifts manually, the linac operators kept the beam energy within a $\pm$1\% range around 28.5~GeV during the run.

\begin{figure}[t!]
  \begin{center}
     \includegraphics[width=0.7\textwidth]{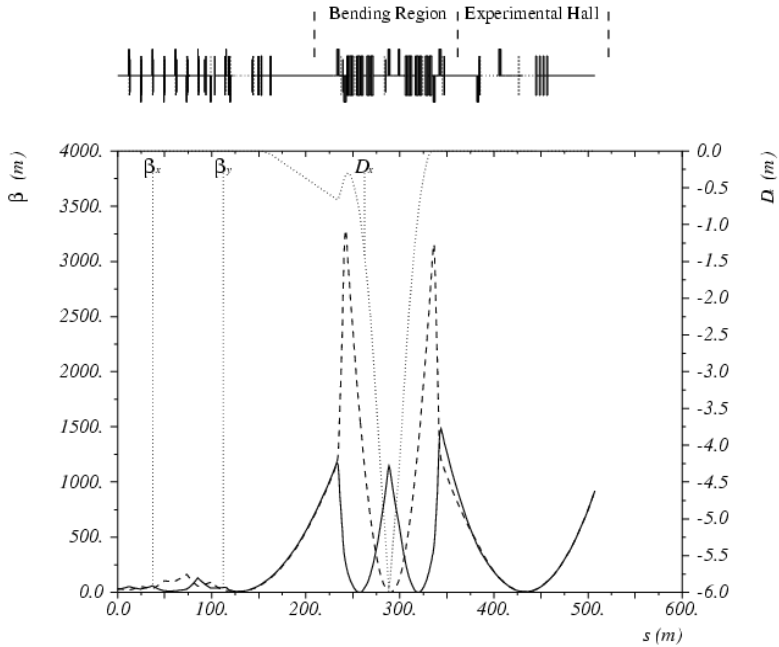}
     \caption[Magnetic chicane in ESA]{Beta functions ($\beta_x$, $\beta_y$) and horisontal dispersion ($D_x$) in the A-line and ESA beamline.}
    \label{fig:esaoptics}
  \end{center}
\end{figure}

Figure \ref{fig:esaoptics} shows the horisontal and vertical beta functions as well as the horisontal dispersion throughout the A-line and ESA. The maximum dispersion and horisontal waist are at the location of the high power momentum slits. The dispersion is then minimised throughout the ESA experimental area. Detailed information on the optics studies in the ESA can be found in \cite{ref:optics1,ref:optics2}

\begin{figure}[t!]
  \begin{center}
     \includegraphics[width=\textwidth]{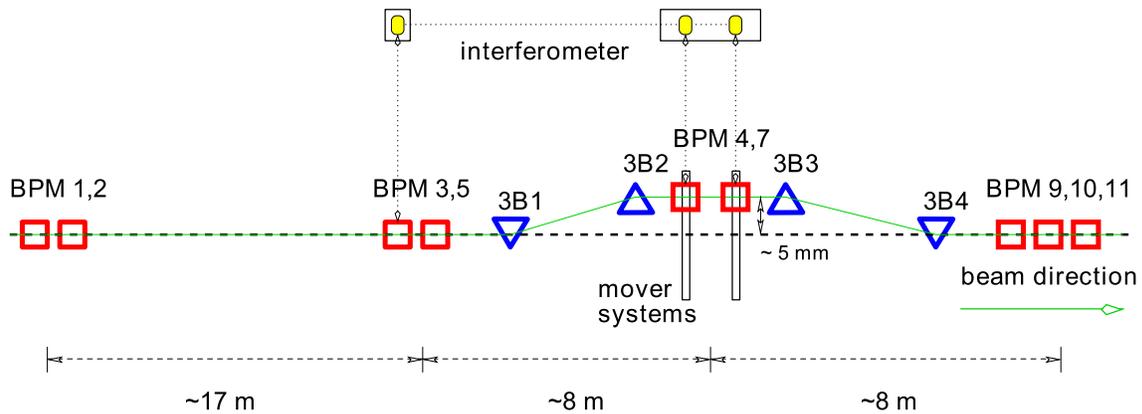}
     \caption[Magnetic chicane in ESA]{Schematic representation of the
       prototype spectrometer in ESA.}
    \label{fig:ESA_beamline_2007}
  \end{center} 
\end{figure}

The setup, as schematically shown in figure~\ref{fig:ESA_beamline_2007}, 
includes four bending magnets denoted as 3B1, 3B2, 3B3 and 3B4, 
forming a chicane in the horizontal plane and high-precision cavity BPMs upstream,
downstream and in between the dipole magnets. Two of them (BPMs 4 and 7) in the middle
of the chicane were instrumented with precision movers. When the magnets were turned on, these BPMs were mechanically moved to ensure the beam offset fits the dynamic range of the BPM electronics. These movers were also used for position calibrations. 
Horizontal positions of three BPMs (3, 4 and 7) were monitored
with a Zygo interferometer~\cite{ref:zygo}.

\begin{figure}[!ht]\centering
  \subfigure{\epsfig{file=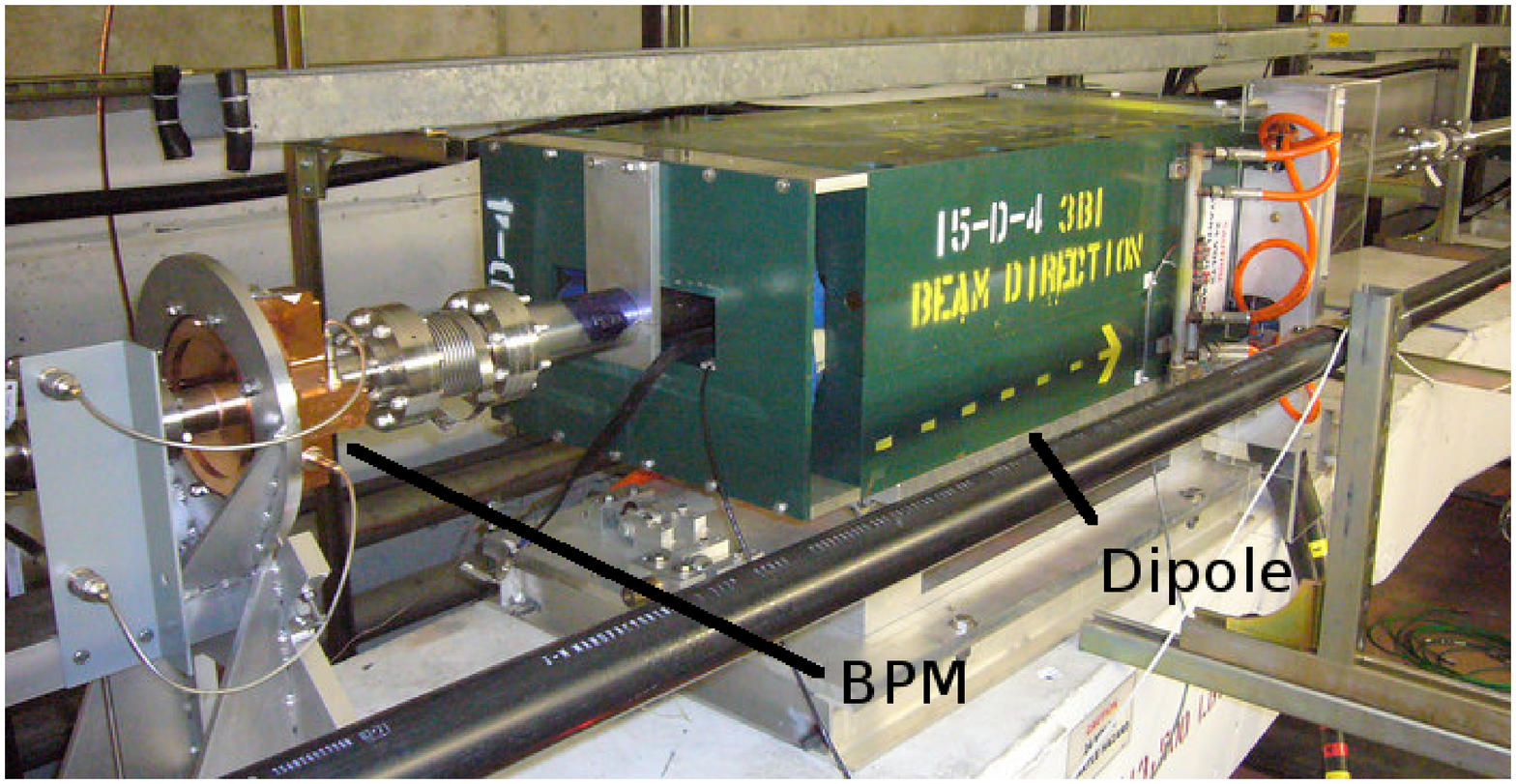,width=0.494\textwidth}}
  \hspace*{0.2cm}
  \subfigure{\epsfig{file=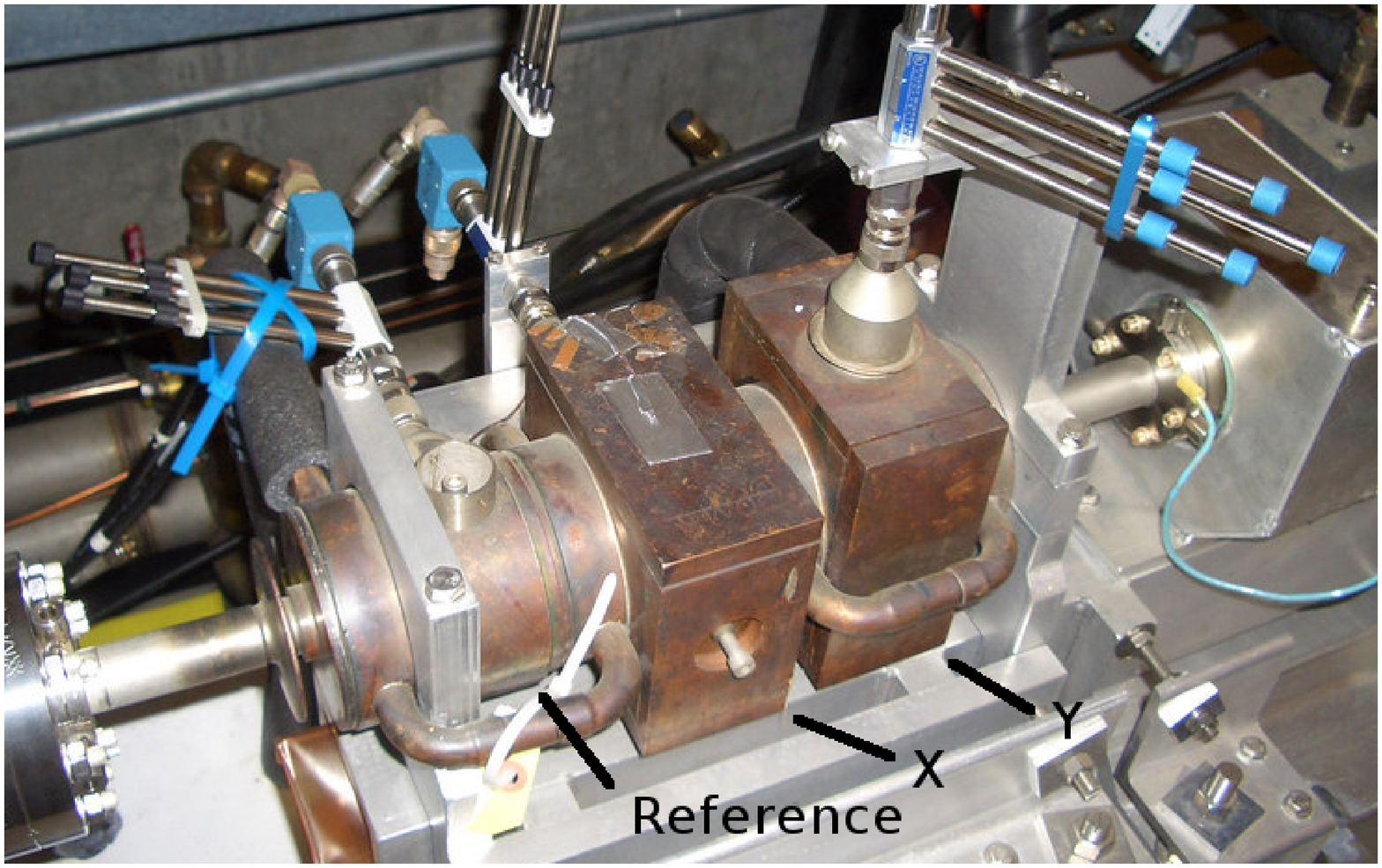,width=0.406\textwidth}}
  \caption{ One of the SLAC ILC type BPMs (BPM~5) followed by the first magnet of the chicane (3B1) in the ESA beamline (left), one of the SLAC type BPMs using rectangular cavities (BPM~9) (right). }
\label{fig:magnetbpm}
\end{figure}

The 10D37 magnets from the old SPEAR injection beamline, refurbished for the use in the chicane, are 37" long,
10" wide on the pole faces and have a 3" gap (approximately 94, 25 and 8~cm respectively), the first magnet of the chicane can be seen in figure~\ref{fig:magnetbpm}.
They were run in series from a single power supply 
to minimise relative drifts. The magnets were studied during a set of measurements in the SLAC Magnet Measurement Laboratory. Magnetic field maps 
of the vertical field component $B_y$ were taken using NMR and Hall probes, 
while each $\int{B \, \mathrm{d}l}$ was measured using a flip coil, which was calibrated against 
a moving wire system. Stability and reproducibility were at the focus of these measurements. 
Details of the field measurements can be found in \cite{ref:Viti,ref:PAC07,ref:Kostromin}.

\begin{figure}[h!]
\begin{center}
\epsfig{file=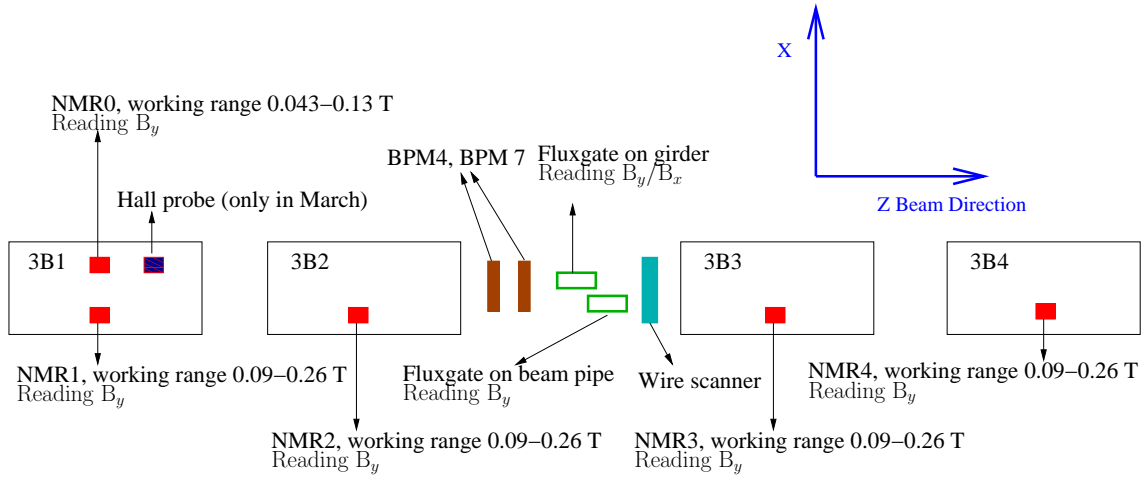,width=\textwidth}
\caption{\label{fig:nmrScheme} Magnetic field diagnostics in the spectrometer chicane.}
\end{center}
\end{figure}

In situ at ESA, two NMR probes with different, but overlapping working ranges
 and initially also one Hall probe were installed in the first magnet 3B1, 
while one NMR probe was positioned in each of the other three magnets,
so that field integral values could be monitored. 
In the test data runs, the nominal magnetic field integral was set at 0.117~T$\cdot$m, 
which corresponds to a current of 150~A. The stray field outside the magnets in the middle of the chicane
was monitored using two low-field fluxgate magnetometers. 
One was placed on the girder to obtain the horizontal ($x$) and vertical ($y$) 
field components and the other
on the beam pipe measuring the $y$-component only. Properties of the probes
and the fluxgate monitors are summarised in figure~\ref{fig:nmrScheme}.

The readout unit for the NMR probes provided one internally-averaged reading every 2.5~s. The probes were multiplexed, sharing the same readout. Typically 9 readings were obtained for each probe before switching to the next probe, totalling an observation time of about 20~s. The gap between observations, while other probes were read out, was about one minute, while an energy scan took about 3 minutes at 10~Hz beam repetition frequency. Therefore, only slow (compared to the data rate) variations of the magnetic field could be tracked reliably.



In order to measure the beam orbit, 8 cavity BPMs, all operating in the RF S-band, were installed. Three of them were SLAC prototype ILC BPMs (3, 4, 5)
using cylindrical cavities with $x$- and $y$-waveguides for the dipole mode coupling
and monopole mode suppression (figure~\ref{fig:magnetbpm}, left). Each of the five SLAC BPMs (A-line-type BPMs 1 and 2, and linac-type BPMs 9, 10, and 11)
consists of three cavities: two rectangular cavities for $x$ and $y$ separately to avoid $x$-$y$ couplings,
and one cylindrical cavity to provide charge and phase information \cite{ref:slac} (figure~\ref{fig:magnetbpm}, right).
BPM~7 was a dedicated ILC prototype designed and manufactured in the UK
for the use in the spectrometer. Unfortunately, this monitor could not be used
in the analysis due to manufacturing problems~\cite{ref:ukbpm}. 
Micrometre level resolution was measured for BPMs 1 and 2, while BPMs 3, 4, 5, 9, 10 and 11 demonstrated a resolution below 1~$\mu$m. Details on the performance of the BPM system and the A-line configuration
can be found in~\cite{ref:BPM}.



BPMs 12 and 24 are placed in the bending arc region of the A-line, where horizontal dispersion reaches about 0.5~m. For our experiment they were instrumented
with the same high-sensitivity electronics as all other BPMs in the ESA beamline, so that the energy measurements in the A-line and in the chicane could be performed simultaneously and cross-checked against each other.




\section{Performance of the prototype spectrometer}

\subsection{Reconstruction of the beam orbit in the middle of the chicane}
\label{txt:bpm4}

As the chicane magnets bend the beam in the $x$-direction, 
we are mainly interested in the horizontal beam position and angle, and, 
unless specified otherwise, we refer to the
 $x$-coordinate throughout this section.

In our system, signals generated by the BPMs were digitised and stored in data files for each event, i.e. for each beam trigger. They are digitally demodulated in the analysis~\cite{ref:BPM}. A complex digital local oscillator signal allows decoding of both the amplitude and the phase of the signal's phasor along the waveform. 
Sampled at a point close to the peak and normalised by the phasor from the reference cavity, 
the converted waveforms give the real, in-phase (I), value and the imaginary, quadrature (Q), value, 
which contain the information on the beam offset as well as the inclination.

The offset of the beam trajectory in the middle of the chicane
has to be measured with respect to the nominal orbit position reconstructed
using BPMs outside of the chicane.
In order to form a prediction of the beam position at the BPM~4 location 
we took data with zero current in the magnets and selected a ``quiet period'', when neither the beam nor the hardware settings were altered. 
We then correlated the I and Q readings of BPM~4 with the data from other BPMs. Forming the prediction can be visualised as continuing the beam trajectory line connecting the points measured by other BPMs up to BPM~4 location. The best set of linear correlation coefficients minimises the offset between that line and the measured points for the majority of the beam passes.

Data from a run with magnets on could also be used for relative measurements 
 and would result in a better prediction, however, due to the
residual dispersion in the beamline, beam positions before
and in the middle of the chicane are correlated.
 Hence, only data from a run with magnets off were used.

BPMs 9, 10 and 11 were not used for the prediction because, when magnets are on, the impact of the chicane on the beam orbit is not fully compensated, and the beam offset in these BPMs is energy-correlated.

Due to alignment errors, there is also a correlation between the vertical beam position 
and angle before the chicane and the horizontal beam position and angle in the mid-chicane.
Therefore, both $x$ and $y$ readings from the BPMs upstream of the chicane
(x1, x2, x3, x5, y1, y2, y3 and y5) were used in the analysis.

In order to reconstruct the beam orbit in the mid-chicane,
the I and Q values from BPM~4 are correlated to the I and Q
values from the upstream BPMs. This means solving an overdetermined set of linear equations:

\begin{equation}
\mathbf{Ax}=\mathbf{b},
\label{lineq}
\end{equation}

\noindent where $\mathbf{A}$ is a matrix containing the readings of the $n$ selected BPMs for $m$ beam pulses (we used several thousands), and an additional unity column for the offset:

\[
 \mathbf{A} =
 \begin{pmatrix}
  I_{1,1} & I_{2,1} & \cdots & I_{n,1} & Q_{1,1} & Q_{2,1} & \cdots & Q_{n,1} & 1      \\
  I_{1,2} & I_{2,2} & \cdots & I_{n,2} & Q_{1,2} & Q_{2,2} & \cdots & Q_{n,2} & 1      \\
  \vdots  & \vdots  & \ddots & \vdots  & \vdots  & \vdots  & \ddots & \vdots  & \vdots \\
  I_{1,m} & I_{2,m} & \cdots & I_{n,m} & Q_{1,m} & Q_{2,m} & \cdots & Q_{n,m} & 1      \\
 \end{pmatrix},
\]

\noindent and $\mathbf{b}$ is the vector of either I or Q readings of BPM~4:

\[
 \mathbf{b}_I =
 \begin{pmatrix}
  I_{\mathrm{BPM4},1} \\
  I_{\mathrm{BPM4},2} \\
  \vdots              \\
  I_{\mathrm{BPM4},m} \\
 \end{pmatrix},
 \hspace{3mm}
 \mathbf{b}_Q =
 \begin{pmatrix}
  Q_{\mathrm{BPM4},1} \\
  Q_{\mathrm{BPM4},2} \\
  \vdots              \\
  Q_{\mathrm{BPM4},m} \\
 \end{pmatrix}.
\]

We applied the Singular Value Decomposition (SVD) method \cite{ref:svd} to solve these equations. From the SVD

\begin{equation}
\begingroup
 \mathbf{A} = \mathbf{U}\cdot
 \begin{pmatrix}
  w_1 &     &        &          \\
      & w_2 &        &          \\
      &     & \ddots &          \\
      &     &        & w_{2n+1} \\
 \end{pmatrix}
\cdot \mathbf{V}^T,
\endgroup
\end{equation}

\noindent and the solutions of eq.~\ref{lineq} can be found as:

\begin{subequations}
\begin{align}
\mathbf{x_I} &= \mathbf{V}\cdot diag(1/w_j)\cdot \mathbf{U}^T\cdot\mathbf{b_I},\\
\mathbf{x_Q} &= \mathbf{V}\cdot diag(1/w_j)\cdot \mathbf{U}^T\cdot\mathbf{b_Q}.
\end{align}
\end{subequations}

\noindent $\mathbf{x_I}$ and $\mathbf{x_Q}$ are vectors of coefficients, which relate
the Is and Qs of all selected BPMs to those of BPM~4 so that a prediction can be made:

\begin{subequations}
\begin{align}
I_{\mathrm{BPM}4,pred} &= \begin{pmatrix}I_1 & I_2 & \hdots & I_n & Q_1 & Q_2 & \hdots & Q_n & 1 \end{pmatrix}\cdot\mathbf{x}_I  \hspace{1mm},\\
Q_{\mathrm{BPM}4,pred} &= \begin{pmatrix}I_1 & I_2 & \hdots & I_n & Q_1 & Q_2 & \hdots & Q_n & 1 \end{pmatrix}\cdot\mathbf{x}_Q  \hspace{1mm}.
\end{align}
\end{subequations}



\begin{figure}[h!]
\begin{center}
\epsfig{file=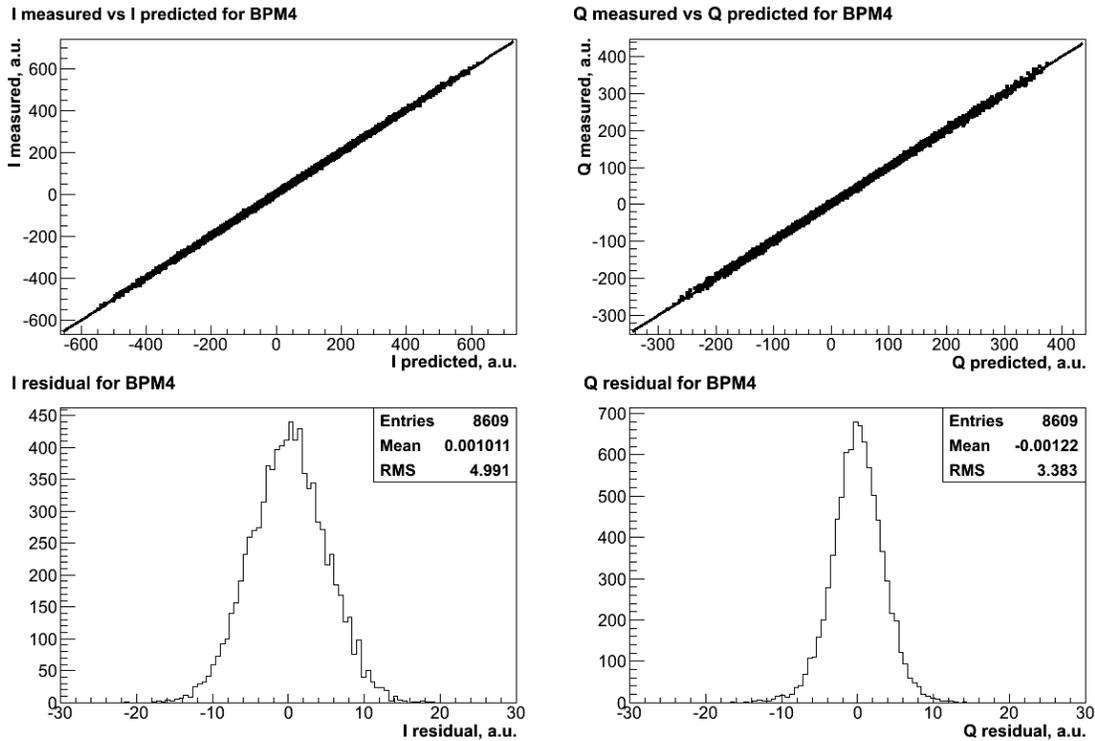,width=\textwidth}
\caption{\label{fig:bpm4residuals} BPM~4 readings predicted from
  other BPMs in the beamline: I predicted vs I measured
  (top left), Q predicted vs Q measured (top right), I 
  residual (bottom left), Q residual (bottom right).}
\end{center}
\end{figure}

The difference between the predicted and the measured values is the residual. 
In our case, the RMS residual is the precision of the orbit prediction
and the resolution of BPM~4 added in quadrature. It sets the limit on the
spectrometer resolution. The measured and predicted values for I and Q are plotted 
against each other in figure~\ref{fig:bpm4residuals}.
The points in these plots lie around the $y=x$ solid lines,
which means the prediction works correctly. The histograms in the bottom part
of figure~\ref{fig:bpm4residuals} show the residuals,
for both the I and Q values.

\begin{figure}[h!]
  \begin{center}
    \epsfig{file=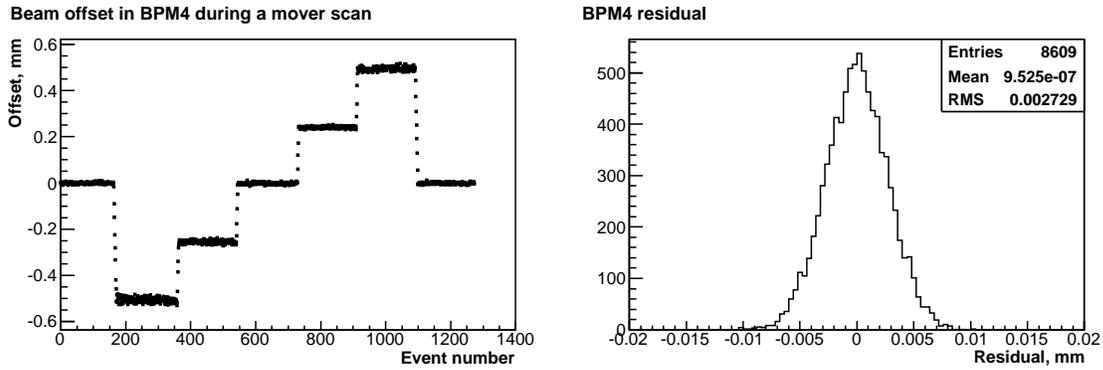,width=\textwidth}
    \caption{\label{fig:bpm4check} BPM~4 position for a horizontal mover scan (left),
    BPM~4 residual during a quiet period (right).}
  \end{center}
\end{figure}

It is clear that the I and Q residuals for BPM~4 are small compared to
the average I and Q values, but the results in figure~\ref{fig:bpm4residuals}
are still hard to interpret quantitatively. In order to set the scale we used
the mover scan data. During the mover scan BPM~4 was
moved in 0.25~mm steps from $-$0.5 to $+$0.5~mm with respect to the nominal position. The precision of the mover system is about 10~$\mu$m, but the moves can also be observed by the interferometer with a sub-micrometre precision.
Figure~\ref{fig:bpm4check} shows the scan data as well as the position
residual, which was calculated for the data used in the SVD computations
above. A position residual of 2.73~$\mu$m was determined, 
which is close to the estimate in~\cite{ref:Viti} (2.3~$\mu$m).

The residual is larger than our earlier published value \cite{ref:BPM}, which was close to 1~$\mu$m. This is due to the movement of BPM~4 from its original location between BPMs~3 and~5 to the middle of the chicane and exclusion of BPMs 9, 10 and 11 from this analysis. Therefore, BPM~4, which was previously in the ``centre of gravity'', here is at the edge of the BPM system. Clearly, the precision of the orbit reconstruction at BPM~4 was affected.

Together with the 5~mm nominal beam offset in the middle 
of the chicane for magnets on, the 2.73~$\mu$m precision of the BPM system sets an energy resolution limit of ${5.5\cdot 10^{-4}}$ for our spectrometer prototype.




\subsection{Estimate of the beam energy and scale correction}
\label{txt:absEne}

The I and Q readings predicted for BPM~4 by all other BPMs can be
subtracted from the measured values and, when the magnets are on, provide
information on how the beam trajectory changes with the energy.

\begin{figure}[ht!]
\begin{center}
\epsfig{file=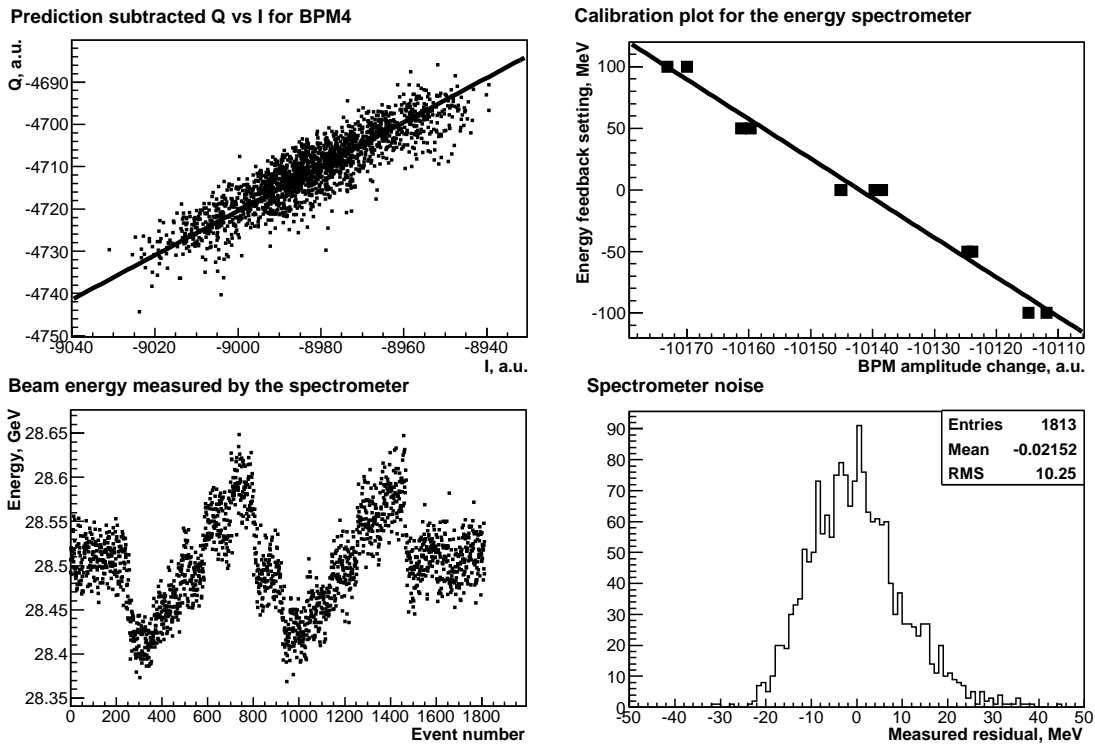,width=\textwidth}
\caption{\label{fig:absEne} Beam energy measurements:
 prediction subtracted Q vs I for BPM~4 (offset by $Q_0$ and $I_0$ to take into account the 5.0034~mm move), with a fit to the data shown (top left), energy calibration plot for the spectrometer (top right), beam energy measured during the scan (bottom left), spectrometer noise measured off the energy line (bottom right).}
\end{center}
\end{figure}

When turning the magnets on, we also moved BPM~4 by 5~mm in order to keep
the beam centred. This movement was observed by the
 Zygo interferometer. 
According to the interferometer, BPM~4 moved by 5.0034~mm between our selected runs with magnets on and magnets off.
Using the IQ rotation and scale 
from the mover scan, we can predict the changes of the I and Q values of BPM~4.
This results in offsets of
$I_0 = -8784$ and $Q_0 = -4605$, 
which were added to the I and Q values from the energy scan after the predictions had been subtracted (figure~\ref{fig:absEne}, top left).

Although a small inclination of the beam orbit is introduced along with the offset in the middle of the chicane due to small differences between the magnets, the measured points still lie on a straight line in the I-Q plane as both the offset and inclination scale with the energy. Fitting the measured data to a straight line going through the origin, we obtain the IQ rotation of this ``energy line''. Energy readings for each point are then calculated as a projection onto the energy line.

In order to compute the energy scale, individual readings are averaged for each step of the energy scan and then fitted to a straight line (figure~\ref{fig:absEne}, top right). The slope of this line gives the energy scale and the offset -- the measured nominal energy.
This procedure results in a beam energy of about 32.6~GeV, while, as mentioned above, it was kept within $\pm$1\% off 28.5~GeV during the run. Although the fit may contribute up to 1.4~GeV uncertainty, introduced by the drifts during the energy scan, the difference is mainly due to the scale of the energy feedback, which was not re-calibrated for the run.


Introducing the values for the total beam offset ${x = 5.117}$~mm, distance between the magnets ${L = 4.014}$~m, and magnetic field integral $\int B \mathrm{d}l = 0.117$~T$\cdot$m into equation (\ref{eqn:energy}) results in a value lower than expected, 27.5~GeV. Nevertheless, this estimate confirms that the beam energy was not as high as measured using the uncorrected energy feedback scale. As measuring the absolute beam energy is out of the scope of this study, and some systematic offsets may contribute to $E_b$, we assume a nominal beam energy of 28.5~GeV in this article.

The ratio 28.5/32.6 gives a correction factor of 0.87, meaning that the energy scan was actually performed in a range of $\pm$87~MeV instead of requested $\pm$100~MeV, and the energy scale factor must be corrected accordingly.

The energy measured by BPM~4 during the scan is shown in figure~\ref{fig:absEne}, bottom left. Peak fluctuations are less or comparable with the energy scan step size of~50~MeV, so a resolution better than 25~MeV can be expected. In the following we use the data from the energy BPMs in order to separate the energy fluctuations from noise, and include additional data acquired with the setup, such as interferometer and NMR readings, to refine the measurement and estimate the resolution of the spectrometer.

The last plot in figure~\ref{fig:absEne} (bottom right) shows the distribution of the offsets of the measured points from the fitted line. The RMS of the distribution is 10~MeV, or 8.7~MeV (${3.1\cdot10^{-4}}$) taking into account the scale correction. This value reflects the noise performance of the BPM system since the energy- and position-induced changes act along the energy line (the incline, although not always negligible, is very small). However, it does not include the effect of the magnetic field, beam position fluctuations and associated non-linearities. Indeed, the resolution estimate of ${5.5\cdot 10^{-4}}$ obtained using position data (see section~\ref{txt:bpm4}) is larger.

\subsection{Resolution of the energy BPMs}



We could only perform a relative energy measurement with BPMs~12 and~24, as the field of the bending magnets in the A-line could not be turned off. However, we were still able to calibrate the energy BPMs using the energy scan data and taking into account the energy feedback scale correction.

Similarly to spectrometer data, we measured the RMS residual between the fitted energy line and the measured points for the energy BPMs 12 and~24.
The measured noise is equivalent to 0.36~MeV for BPM~12 and 2.0~MeV for BPM~24,
or ${1.3\cdot 10^{-5}}$ and ${7.0\cdot 10^{-5}}$ respectively, at the nominal beam energy of 28.5~GeV.
The values are different because BPM~12 
had an additional 20~dB amplifier installed in its electronics chain 
in order to compensate for cable losses. As a consequence, this BPM's sensitivity was improved and the impact of the noise 
and granularity introduced by the digitisers was reduced.

\begin{figure}[h!]
\begin{center}
\epsfig{file=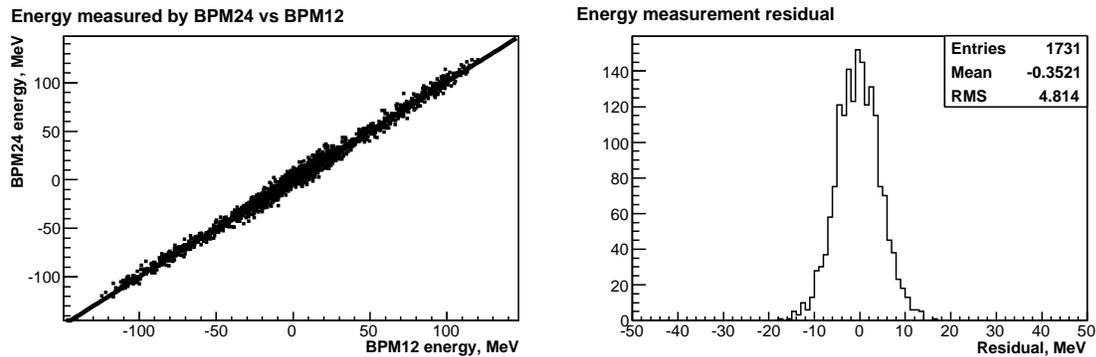,width=\textwidth}
\caption{\label{fig:res12vs24} Comparison of BPMs 12 and 24: 
  BPM~24 vs BPM~12 energy measurement (left), 
  residual between BPM 12 and 24 measurements (right).}
\end{center}
\end{figure}

Again, these estimates only take into account the noise in the BPMs, but not other effects such as the beam jitter and magnetic fields changes.
In figure~\ref{fig:res12vs24} we compare the energy readings of BPMs 12 and~24 
after the energy calibration. An RMS residual of 4.8~MeV
(${1.7\cdot 10^{-4}}$) was found, which is about twice bigger than the noise measurements combined in quadrature. This means that the resolution
of the energy measurements of BPMs 12 and 24 is, in fact, not limited by the BPM noise alone.
Nevertheless, BPMs 12 and 24 still allow energy fluctuations to be measured to better than ${1.7\cdot 10^{-4}}$, which is well below the expected spectrometer resolution.

\subsection{Dipole magnets}

An essential prerequisite for the operation of the spectrometer in a Linear Collider is that the beam
position downstream of the chicane is not energy-dependent, and the upstream beam path 
is restored downstream. In other words, the chicane has to be symmetric. In a 4-magnet chicane it is also beneficial to match the magnets in each pair producing a parallel translation of the beam (a ``dogleg''), so that the inclination of the orbit with respect to the original is kept to a minimum. 

\begin{figure}[ht!]
  \begin{center}
    \includegraphics[width=\textwidth]{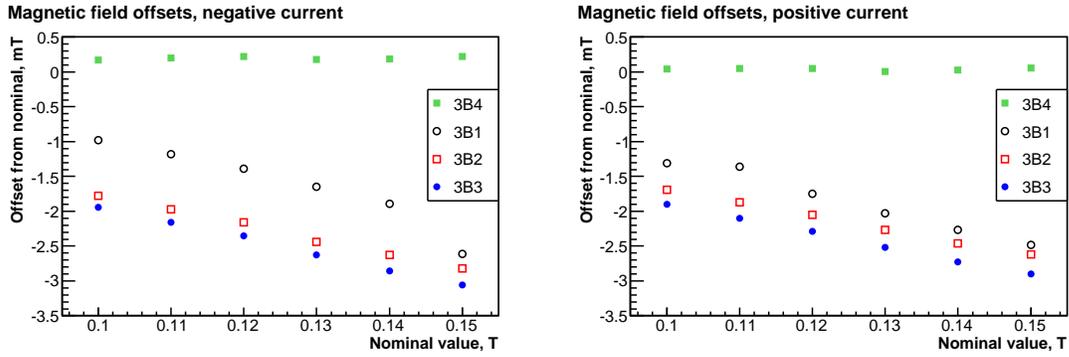}
    \caption{Offsets between the measured and nominal magnetic fields as
      a function of the nominal value of the four magnets in ESA:
      Negative current (left); Positive current (right).}
    \label{fig:MeasurvsNom}
  \end{center} 
\end{figure}

Magnetic field measurements were performed in March 2007. Some results are shown in
figure~\ref{fig:MeasurvsNom}. Here, the differences between the measured and
nominal magnetic fields are plotted as a function of the nominal value for
both negative and positive polarities.

During these measurements the field of the magnet 3B1 was monitored with a Hall probe, whereas for the other magnets NMR probes were used.
As can be seen, 3B1, 3B2 and 3B3 follow the same trend,
with a difference of a few tenths of a mT between 3B2 and 3B3, while 3B1
differs by about 1 mT.
 Offsets between these magnets can be explained by the individual history and core composition of each (see \cite{ref:Viti} for details).
 3B4 shows a different and much more consistent behaviour, because
only for this magnet a more accurate relation between the current
and the field (as given in \cite{ref:Viti}) was determined and
used for the field settings. Unfortunately, analogous measurements could not be performed for the other magnets due to time constraints. 

For stability, the magnets were powered by a single supply in ESA, therefore, the differences could not be compensated for. As a result, the trajectory of the
beam had a small inclination in the middle of the chicane and was not fully restored downstream of the chicane, and energy changes were converted into position variations in BPMs 9, 10 and 11.





\begin{figure}[ht!]
  \begin{center}
    \epsfig{file=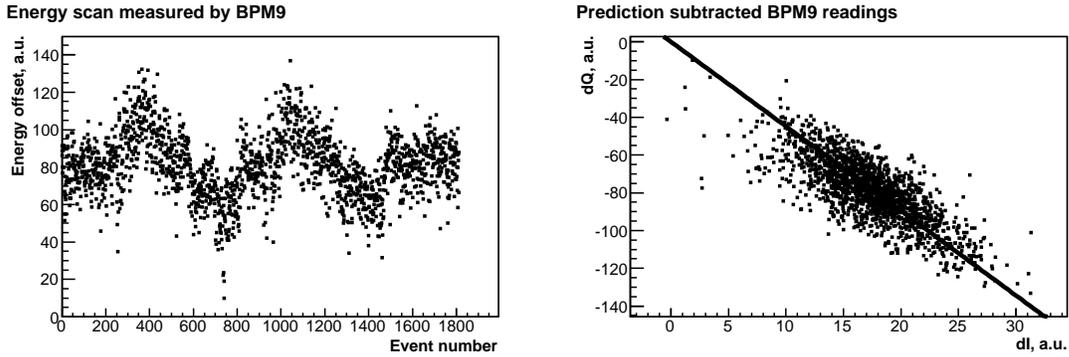,width=\textwidth}
    \caption{\label{fig:bpm9} Energy measured by BPM~9 during the scan (left), 
    IQ plot of the measured BPM~9 readings with the predicted readings subtracted (right).
    The fitted line shows the IQ rotation of the energy measurements.}
  \end{center}
\end{figure}

Using the data from the upstream BPMs the nominal beam position
in the downstream BPMs can be predicted. Considering, for example, BPM~9 measurements after
subtraction of the upstream BPMs prediction, we can recognise the step-like behaviour of the energy during the scan (figure~\ref{fig:bpm9}).
Note that, although the net integral field applied to the beam by the chicane
is very small, 
BPM~9 is still able to resolve the energy changes due to its high resolution.



\subsection{Energy resolution of the spectrometer}
\label{txt:eneRes}

The energy measured by the spectrometer 
can also be predicted by the energy BPMs 12 and 24. The residual, besides
the resolutions of each BPM, depends on the fluctuations of the
magnetic fields, mechanical vibrations, as well as drifts and other
systematic effects and non-linearities.

\begin{figure}[h!]
\begin{center}
\epsfig{file=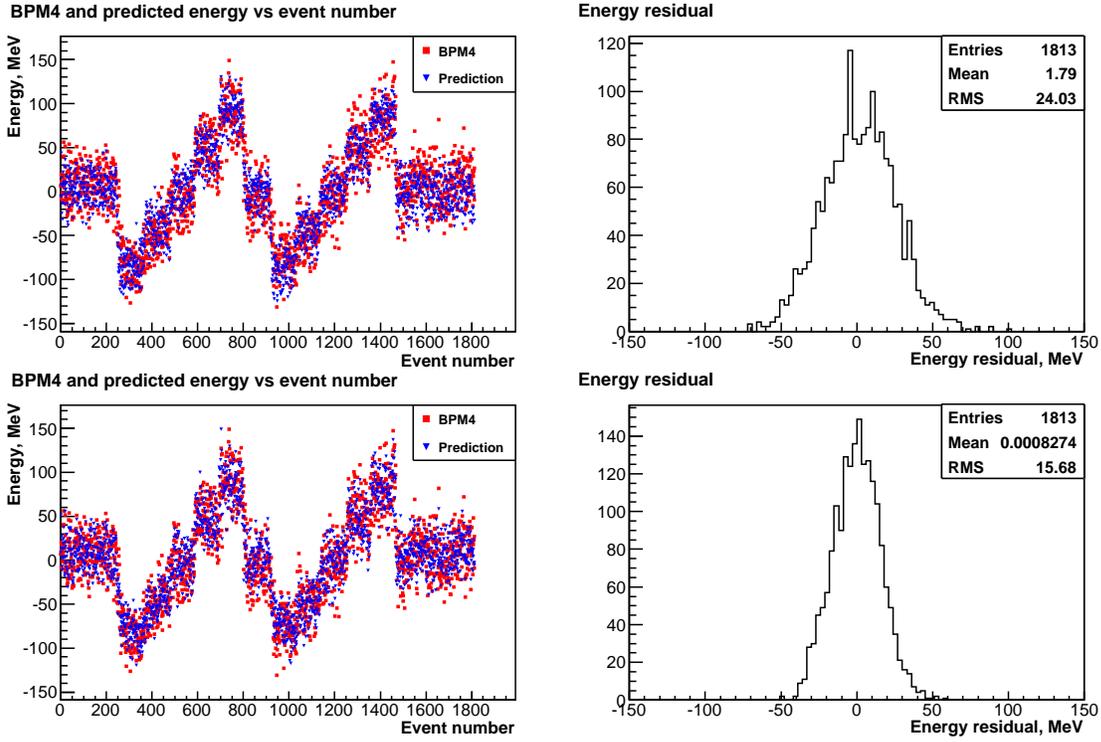,width=\textwidth}
\caption{\label{fig:eneRes} Energy resolution measurement: energy
  measured by BPM~12 and BPM~4 (top left), residual between BPM~12 and
  BPM~4 readings (top right), energy measurement predicted by BPMs 12,
  24 and additional parameters and BPM~4 reading (bottom left), residual
  between the prediction and BPM~4 reading (bottom right).}
\end{center}
\end{figure}

We first compare the relative energy measured by BPM~4 with the measurements
of BPM~12 (figure~\ref{fig:eneRes}, top). This results in a
resolution of 24~MeV or ${8.4\cdot 10^{-4}}$. As this is worse
than the precision of the orbit reconstruction, we decided to look for
correlations using additional data and applying the SVD method by starting again from
BPM~12 and then adding more data in the matrix to better reconstruct the spectrometer measurements and understand the systematics.

Each time we added another parameter to the matrix, we re-calculated the SVD coefficients from the energy scan data and then applied them to the data from the quiet period. For both data sets we calculated the RMS energy residual (table~\ref{tab:eneResiduals}). Note that this time when we compare BPM~4 and BPM~12 measurements the scale is corrected by the SVD for a better match, which results in a lower residual.

Where the residual is improved for both the energy scan and quiet period, we can conclude that the uncertainty associated with the included parameter is reduced. We also estimate that uncertainty ($\Delta\sigma/\sigma$) subtracting the residuals ($r$) in quadrature and normalising the result by the nominal energy: $\Delta\sigma/\sigma = \sqrt( r_{previous}^2 - r_{current}^2 )/E_b$. These estimates are also shown in table~\ref{tab:eneResiduals}.

The biggest residual reduction is observed when the data from BPMs 9, 10 and 11 are included in the computation. As we know, these BPMs are sensitive to the energy. In addition, these BPMs outperform the rest of the BPMs in the beamline by almost an order of magnitude in terms of resolution \cite{ref:BPM}. For that reason, even though the net field of the chicane is small, they form another spectrometer arm with a comparable resolution.


Some further improvement is also noted when the bunch charge $q$, as measured by one of the reference cavities, is taken into account,
even though all the BPM data were normalised by the charge. This is best explained
by the fact that BPMs 12 and 24, although very sensitive to energy changes, 
were not centred in their operating ranges, and were running close to saturation. As a consequence, non-linearities could be introduced.

Ultimately, in order to achieve an
energy resolution approaching $10^{-4}$, one has to monitor the
relative motion of the BPMs in the beamline.  An interferometer,
once well tuned, seems to be a reliable, fast and high precision tool. 
Since the mechanical vibrations observed were in the order of a few
hundred nanometres, the Zygo interferometer in our setup only
provided a moderate improvement to the energy measurement.

Our system did not provide bunch-to-bunch magnetic field measurements, therefore only interpolated field data could be used. Inclusion of such data in the analysis did not provide a consistent improvement, but the field data themselves suggest that relatively fast fluctuations of the magnetic field take place.

The final
result of these investigations is shown in the bottom part of figure~\ref{fig:eneRes}. 
The resolution was measured
to be 15.7~MeV (${5.5\cdot 10^{-4}}$) for an energy scan and 14.6~MeV (${5.1\cdot 10^{-4}}$)
for a quiet period. These numbers are in a good agreement with the estimate for the precision
of the orbit reconstruction of ${5.5\cdot 10^{-4}}$, which means that the weighting of different systematics has been performed correctly.

\begin{table}
\begin{center}
  \caption{\label{tab:eneResiduals} Energy residuals calculated for
    BPM~4 including additional parameters. $\Delta \sigma/\sigma$ is the uncertainty due to the added parameters calculated as two consequent residuals subtracted in quadrature and normalised by the nominal beam energy. }
\vspace*{0.1cm}
\begin{tabular}{|l||c|c|c|c|} \hline
\multirow{2}{*}{Data included} & \multicolumn{2}{|c|}{Residual, MeV} & \multicolumn{2}{|c|}{$\Delta \sigma/\sigma$, $\times 10^{-4}$} \\ \cline{2-5}
                        & energy & quiet  & energy & quiet  \\
                        & scan   & period & scan   & period \\ \hline \hline
BPM 12                  & 23.45  & 21.53  & --     & --     \\ \hline
BPMs 12, 24             & 23.08  & 21.64  & 1.5    & 0.8 (up) \\ \hline
BPMs 12, 24 and NMR     & 22.67  & 22.62  & 1.5    & 2.3 (up) \\ \hline
BPMs 12, 24, NMR        & 22.67  & 22.62  & --     & --     \\
and fluxgate            &        &        &        &        \\ \hline
BPMs 12, 24, charge (q),& 20.52  & 19.68  & 3.4    & 3.9    \\
NMR and fluxgate        &        &        &        &        \\ \hline
BPMs 12, 24, 9, 10, 11, & 15.86  & 15.26  & 4.6    & 4.4    \\
q, NMR and fluxgate     &        &        &        &        \\ \hline
BPMs 12, 24, 9, 10, 11, & 15.68  & 14.60  & 0.8    & 1.6    \\
q, NMR, fluxgate and    &        &        &        &        \\ 
interferometer          &        &        &        &        \\ \hline
\end{tabular}
\end{center}
\end{table}





\subsection{X to Y coupling}

Even though the spectrometer chicane operates in the horizontal plane, 
the energy scan is also traced in the vertical plane. Firstly, 
alignment errors generate a small bend in the vertical direction and,
secondly, internal cross-talk between the $x$- and $y$-couplers of the BPMs
create a spurious offset in $y$ due to an offset in $x$ and vice versa.

\begin{figure}[h!]
\begin{center}
\epsfig{file=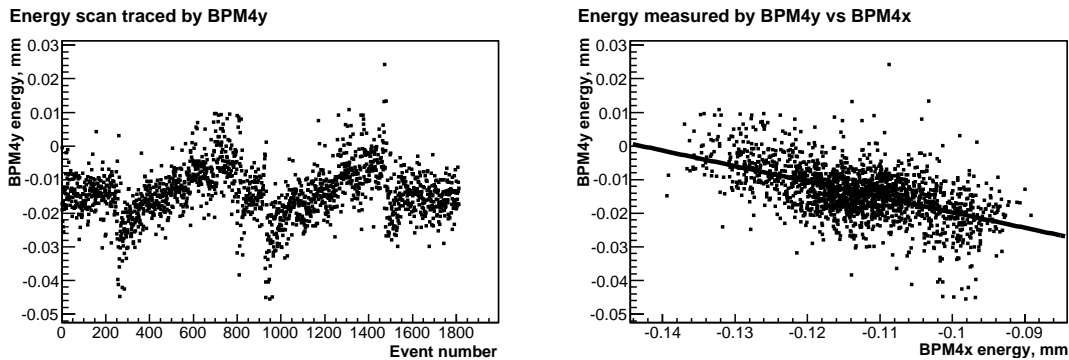,width=\textwidth}
\caption{\label{fig:bpm4y} Effect of the chicane on the vertical beam trajectory: 
 energy scan traced by BPM~4 in $y$ (left), energy data measured by BPM~4 in $y$ vs $x$ (right).
 Position calibration was used to exclude the difference in sensitivities. 
 Hence, the energy is expressed in terms of the offset (mm).}
\end{center}
\end{figure}

In order to estimate the cross-coupling between the $x$ and $y$ coordinates we again consider the energy scan data, this time to predict
the vertical beam position in BPM~4 using the SVD coefficients obtained from the run with magnets off.
Clearly, as seen in figure~\ref{fig:bpm4y} (left), the energy scan is traced 
in the measured $y$ offset.
Due to different sensitivities of the $x$ and $y$ channels in BPM~4,
we used mover scan data in both directions to get the position scales,
which are used to normalise the raw energy. For that reason the energy is
given in terms of mm in figure~\ref{fig:bpm4y}. One should, however, keep in mind
that an energy change generates both a different offset and an inclination
in the mid-chicane.

The plot on the right-hand side in figure~\ref{fig:bpm4y} shows the correlation between the energy measured
in both planes. From the inclination of the line fitting the data points
a rotation of BPM~4 of almost 25$^{\circ}$ is derived, or an $x$-$y$ isolation of about 7.6~dB. Even without tuning, BPMs usually provide an isolation of 20~dB, which means that the cross-talk can not be explained solely by the cross-coupling of the signals. At the same time, the rotation is too large to be caused entirely by the alignment errors. This indicates that both effects take place. For the future, it is therefore important to minimise the cross-talk in the BPMs
and eliminate fake offsets by careful alignment of the spectrometer elements.

\section{Suggestions for future experiments}

Clearly, any improvement of the BPM resolution would have a significant positive impact on both the relative and absolute energy measurement as it reduces the BPM uncertainties contributing to the overall measurement error.

Improvement of the internal $x$-$y$ isolation in the BPMs would also have a positive impact on the energy measurement as the uncertainty introduced by the signal cross-coupled from the orthogonal direction would be smaller. Isolations of 40~dB and higher can be achieved with current designs.

Higher resolution BPMs could also simplify the operation of the spectrometer. For a 1~mm dispersion, a resolution of 100~nm would give a 10$^{-4}$ energy resolution. Currently, a dynamic range of about 80~dB can be achieved with cavity BPMs, which allows 1~mm offsets to be measured with no need to move the BPMs. Hardware improvements and better algorithms to treat the signals saturating the electronics \cite{ref:atfbpms} are expected to expand the dynamic range to 90 and even 100~dB, although with some degradation of resolution at large signal levels. Additional non-linearities can be calibrated out through a wide-range position scan.


Without the need to move the BPMs when the chicane is in operation, the requirements on precision movers for position calibration can be relaxed, although simpler movers are still mandatory for calibrating out non-linearities and alignment. A direct calibration of the spectrometer can be performed by changing the phase of the RF in some accelerating modules, as it was done in our ESA experiment. Another way of calibration is to change the magnetic field by a small but known amount and restore the energy scale from the orbit changes.

Working with I and Q values of the BPMs directly, we show that even a 4-magnet chicane does not generate a pure beam offset in the middle of the chicane because of small differences between the magnets. At the required level of precision the inclination still needs to be taken into account. Furthermore, two magnets contribute to the uncertainty of the energy measurement in a 4-magnet chicane.

\begin{figure}[h!]
\begin{center}
\epsfig{file=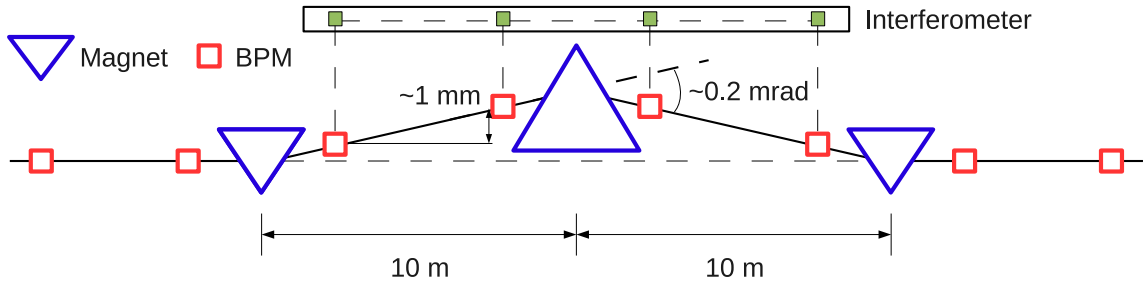,width=\textwidth}
\caption{\label{fig:3magnet} A 3-magnet spectrometer chicane.}
\end{center}
\end{figure}

These arguments suggest a revival of the original 3-magnet chicane design as discussed in \cite{ref:lcdet04-31} and shown in figure~\ref{fig:3magnet}, where the central magnet, the spectrometer magnet, is instrumented with probes and the other two help to preserve the initial beam trajectory. The spectrometer magnet can also be combined of two half-strength magnets, so that all the chicane magnets are identical as they are in a 4-magnet chicane. High-precision BPMs in between the magnets provide information on the bend of the beam, while BPMs upstream of the first magnet predict the default trajectory downstream. In this case, the spectrometer magnet produces a combination of offset and angle in the BPMs downstream, but all measured data should still lie on one line in the I-Q space as in our analysis, see section~\ref{txt:absEne}.

Instrumenting the ancillary magnets and extending the interferometer 
onto the up- and downstream BPMs would provide redundant energy measurement 
at a low increment in cost. While the overall resolution is not expected to become 
improved as the ancillary magnets operate at half of the magnetic field of the spectrometer magnet, some systematic effects can be a priori excluded due to the opposite bend. Also, BPM triplets instead of doublets in between the magnets would also provide redundancy of beam orbit measurements and improve both the precision and accuracy of the spectrometer.

To predict the default trajectory in a 3-magnet spectrometer, the I-Q space of the BPMs can be scanned by changing the beam deflection of the ancillary magnets, while the spectrometer magnet is off.

A precision interferometer will be required to achieve the 10$^{-4}$ or better beam energy uncertainty. This becomes critical for a reduced dispersion as the BPM resolution must be enhanced to 100~nm, since RMS vibrations measured at ESA were about 300~nm for stationary BPMs and approached 1~$\mu$m for BPMs mounted on the movers. The Zygo interferometer fulfils the requirements of the energy spectrometer, hence the vibrations should not present a problem in future installations.

The resolution of the spectrometer also depends on the stability of the magnetic field between the NMR measurements. The time resolution of the NMR probes is in the order of tens of milliseconds, which is sufficient for bunch train averaged measurements in a linear collider, but not for bunch-by-bunch operation. Stabilised low-noise power supplies for the magnets, dedicated readout for each probe (no multiplexing), and combination of NMR and Hall probes will help improve the accuracy of the bunch-by-bunch measurements. Current feedbacks based on the NMR measurements can also help improving the long-term stability of the magnetic fields.

\section{Summary}

The model-independent analysis of the data obtained with the prototype Linear Collider 
spectrometer based on a magnetic chicane achieved a single-bunch resolution of ${5.5\cdot 10^{-4}}$ using a BPM system with a micrometre level precision
of the beam orbit measurements. This value satisfies the requirements for the Linear Collider in most scenarios, and can be improved. Note, that it should not be mistaken for the absolute accuracy, which requires further studies including cross-comparison with an instrument using different physical principle and collision events.

An improved BPM resolution is the key factor 
to enhance the energy resolution. To achieve the $10^{-4}$ level, stabilisation of the magnetic field in the chicane combined with 
fast and reliable field measurements and monitoring of the relative BPM motion in the horizontal plane are also mandatory.

Novel signal processing and analysis techniques allow the BPM resolution to be pushed to the 100~nm level and below, while enhancing the dynamic range of cavity BPMs beyond the current limit of approximately 80~dB, so that large beam offsets can still be measured. This means that the dispersion in the chicane, and hence the beam emittance degradation caused 
by the spectrometer, can be significantly reduced. Further improvements of the BPM resolution and their dynamic range would allow operation of the chicane without BPM movers, eliminating 
associated systematic errors.

Working with uncalibrated in-phase and quadrature BPM readings, one does not have to distinguish between the beam angle and offset changes 
in the middle of a 4-magnet chicane. Both the angle and offset 
follow the energy changes, and the IQ readings produce a straight line in the IQ plane. However, an energy calibration of the whole system may be required in this case. It is also possible to work with calibrated offsets, providing the chicane magnets are closely matched.

For simplicity reasons, a 3-magnet chicane may be a possible configuration. In this configuration, the energy calibration of the chicane becomes necessary. Hence, any reference to a well known physics quantity,
such as the Z-mass, or a complementary method to measure $E_b$, is important
for both the scale corrections of the relative measurements and establishing the offset 
for absolute energy measurements.


\acknowledgments

We would like to thank all SLAC staff who helped with the experiment and machine operation. We would also like to thank our funding bodies, in particular:\\
The Commission of the European Communities under the 6th Framework Programme ``Structuring the European Research Arm,'' contract number RIDS-011899 and the Science and Technology Facilities Council (STFC), LCABD program for funding the UK institutions.\\
The U.S. Department of Energy under contract DE-AC02-76SF00515 for supporting this work at SLAC.\\
The U.S. Department of Energy under contract DE-FG02-03ER41279 for supporting the colleagues at University of California and LBNL.\\
The Research Corporation under contract NSF PHY0529471, and the U.S. Department of Energy under contract DE-FG02-05ER41383 for funding the work in University of Notre Dame.



\end{document}